\documentclass[amsmath,amssymb,aps,twocolumn,superscriptaddress,showkeys,citeautoscript,citeautoscript]{revtex4-2}
\usepackage{graphicx}
\usepackage{xcolor}
\usepackage[colorlinks,linkcolor=blue,citecolor=blue,urlcolor=blue]{hyperref}
\usepackage{amssymb}
\usepackage{comment}
\usepackage{xparse}
\usepackage{ifthen}
\usepackage{dcolumn}
\usepackage[flushleft]{threeparttable}
\usepackage{siunitx}

\newboolean{draft}
\setboolean{draft}{false}
\NewDocumentCommand\todo{m}%
{%
  {\color{blue} (TODO: #1)}%
}

\NewDocumentCommand\change{om}%
{%
  \ifthenelse{\boolean{draft}}{
    \IfNoValueTF{#1}{}%
    {%
      {\color{gray}[#1]}
    }%
    {\color{orange}#2}%
  }%
  {#2}%
}

\def\CB{\mathrm{C}_\mathrm{B}}
\def\CN{\mathrm{C}_\mathrm{N}}
\def\mB{\mu_\mathrm{B}}
\def\mN{\mu_\mathrm{N}}
\def\mC{\mu_\mathrm{C}}
\def\CBCN{\CB\CN}
\def\CBCNt{{(\CBCN)}_{2}}

\begin{document}

\title{Blue quantum emitter in hexagonal boron nitride and carbon chain tetramer: proposition of identification}

\author{Marek Maciaszek}
\email{marek.maciaszek@pw.edu.pl}
\affiliation{Faculty of Physics, Warsaw University of Technology,
  Koszykowa 75, 00--662 Warsaw, Poland}
\affiliation{Center for Physical Sciences and
  Technology (FTMC), Vilnius LT--10257, Lithuania}
\author{Lukas Razinkovas}
\affiliation{Center for Physical Sciences and
  Technology (FTMC), Vilnius LT--10257, Lithuania}
\affiliation{Department of Physics/Centre for Materials Science and Nanotechnology, University of Oslo, P.O. Box 1048, Blindern, Oslo N-0316, Norway}

\begin{abstract}
  Single photon emitters in hexagonal boron nitride offer a gateway to the future of quantum technologies, yet their identification remains challenging and subject to ongoing debate. We demonstrate through \textit{ab initio} calculations that the optical properties of a carbon chain tetramer are in excellent agreement with the characteristics of a blue quantum emitter in hexagonal boron nitride emitting at 435 nm. Its calculated zero-phonon line energy (2.77~eV) and radiative lifetime (1.6~ns) perfectly align with experimental observations. The relatively weak electron--phonon coupling (Huang--Rhys factor of 1.5) indicates intense emission at the zero-phonon line. Despite the absence of an inversion center in the carbon tetramer, we demonstrate that it exhibits a negligible linear Stark effect, consistent with experimental findings. Additionally, our hypothesis explains the experimental observation that the formation of blue emitters is only possible in samples containing numerous ultraviolet emitters, which are likely identical to carbon dimers. \vspace{5em}
\end{abstract}

\maketitle

\section{Introduction}
\medskip

%Hexagonal boron nitride (h-BN) has been used in nanotechnology for many years, yet recent attention towards this material has surged,driven by the discovery of numerous single-photon emitters within h-BN

Hexagonal boron nitride (h-BN) has long been employed in nanotechnology, yet recent years have seen a notable surge in attention toward this material, spurred by the discovery of numerous single-photon emitters within h-BN~\cite{Caldwell}. Its wide bandgap enables it to accommodate single-photon
emitters from the near-infrared to ultraviolet spectrum. Notably, many emitters
demonstrate high brightness and stability even at room temperature, positioning
them as promising candidates for applications in emerging quantum
technologies~\mbox{\cite{Tran,Sajid,Aharonovich_Nanolett}}.\looseness=1

A blue quantum emitter with a zero-phonon line (ZPL) of 2.85~eV (435~nm) in
hexagonal boron nitride was observed in 2019 by Shevitski
et~al.~\cite{Shevitski}. Subsequently, its optical properties have been
extensively explored in numerous studies~\cite{Zhigulin1, Zhigulin2, Fournier,
  Shevitski, Gale}. Its excited state lifetime was determined as 1.8--2.1
ns~\cite{Gale, Fournier}. The quantum efficiency of the blue emitters was
estimated to be at least 20\%, making them bright while exhibiting excellent
photostability. It~was concluded that the defect, being the blue quantum
emitter, has no metastable state, and the suggested electronic structure
comprises two levels within the bandgap: one fully occupied in the lower half
and an empty one in the upper half~\cite{Zhigulin1}. Moreover, it was
demonstrated that the emitter exhibits a very weak Stark effect induced by
out-of-plane fields and a much stronger but almost purely quadratic for in-plane fields~\cite{Zhigulin2}.

\change[% OLD
Blue emitters can be created at chosen locations by electron irradiation
treatment~\cite{Fournier}, and this procedure is efficient only in samples
exhibiting a high concentration of ultraviolet (4.1~eV) emitters~\cite{Gale}. It
was observed that the blue emitters remain stable during annealing only up to
800\si{\celsius}~\cite{Chen_annealing}. After annealing to 1000\si{\celsius},
they disappear, but other emitters appear, which ZPLs belong to the range of
420--480~nm (2.58--2.95~eV).%
]{% NEW
  Blue emitters can be created at chosen locations by electron irradiation
  treatment~\cite{Fournier}, and this procedure is efficient only in samples
  exhibiting a high concentration of ultraviolet (featuring a ZPL of 4.1~eV)
  emitters~\cite{Gale}. These blue emitters remain stable under annealing
  treatments up to 800\si{\celsius}~\cite{Chen_annealing}. Nonetheless,
  annealing at 1000\si{\celsius} leads to the disappearance of the blue emitters and the emergence of other emitting centers. The ZPLs of these new emitters are observed across a spectral range from 420 to 480~nm (2.58 to 2.95~eV).}

\change[Based on the experimental observations, it was concluded that the
permanent transition dipole moment of the blue emitter is zero or very small
(for me it is unclear what does this mean). Consequently, an interstitial defect
including nitrogen or carbon atoms was proposed as the origin of the blue
emission~\cite{Zhigulin2}, and the hypothesis suggesting the negatively charged
nitrogen split interstitial defect as the origin of blue emission was
extensively developed~\cite{Ganyecz}. While many of the calculated properties of
the nitrogen interstitial match well with experimental observations, there are
two problematic points in this assignment: (1)~its calculated excited state
lifetime is 54~ns, and (2)~the correlation between the presence of blue and UV
emitters, commonly identified as carbon dimers, cannot be simply explained.]{
  A minimal linear Stark effect has prompted the hypothesis of nitrogen or
  carbon interstitial defects being responsible for blue
  emission~\cite{Zhigulin2}.
  % \textcolor{red}{As a result [becouse of Stark effect]}, an interstitial defect
  % comprising nitrogen or carbon atoms has been proposed to account for the blue
  % emission~\cite{Zhigulin2}.
  This idea was further elaborated with the theory
  that the negatively charged nitrogen split interstitial defect is the
  source of the blue emission~\cite{Ganyecz}. Although the calculated properties
  of the nitrogen interstitial largely align with experimental data, there are
  two significant issues with this assignment: (1)~the computed radiative lifetime is 54~ns, and (2)~the observed correlation between blue and UV emitters, typically attributed to carbon dimers, cannot be simply explained.}

\begin{figure}
  \centering
  \includegraphics[width=0.4\textwidth]{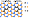}
  \caption{{Cis} and \emph{trans} configurations of carbon tetramers $\CBCNt$ in h-BN.\label{fig:defects}}
  \vspace{-0.5em}
\end{figure}

In the following, we demonstrate that all these experimental observations can be
well explained by the properties of the carbon chain tetramer $\CBCNt$, where
four carbon atoms substitute on four neighbouring sites within the h-BN lattice
(see Fig.~\ref{fig:defects}). Our reasoning is based on four arguments:
\change[% OLD
(1)~the correlation between the presence of blue and UV emitters is easily
explained if both emitters are carbon complexes, (2)~the calculated theoretical
value of ZPL of the carbon tetramer is 2.77~eV, closely matching the
experimental value, (3)~the calculated excited state lifetime is 1.6~ns, in
excellent accordance with the experiment, and (4)~the linear Stark effect of the
carbon chain tetramer is predicted be very small.]%
{%NEW
  (1)~the simultaneous presence of both blue and UV emitters becomes understandable when identifying the 435 nm emitter as a carbon-based complex, (2)~the ZPL
  of the carbon tetramer, calculated to be 2.77~eV, closely matches the experimental value, (3)~the computed radiative lifetime
  of 1.6~ns shows excellent agreement with experimental findings, and (4)~the
  linear Stark effect is predicted to be negligible.
}

The carbon chain tetramer can exist in two geometric configurations:
\textit{cis} or \textit{trans} (see Fig.~\ref{fig:defects}). While calculated
results \change[clearly favor]{favor} the \textit{trans} conformer, which
exhibits better agreement with experimental findings, both conformers will be
analyzed and discussed.

\section{Methodology}

% Spin-polarized density functional theory (DFT) calculations were conducted using the VASP code~\cite{Kresse_vasp}. We employed the projector augmented wave (PAW) method, setting the plane wave energy cutoff to 500~eV~\cite{Blochl}. Our computational model utilized a 288-atom hexagonal supercell, equivalent to a $6\times6\times2$ arrangement of the primitive cell, with Brillouin zone sampling restricted to the $\Gamma$ point. Ionic relaxation proceeded until the forces on all ions were reduced to below 0.01~eV/\AA.

\textit{Ab initio} calculations were performed using spin-polarized density
functional theory (DFT) as implemented in VASP code~\cite{Kresse_vasp}. The
projector augmented wave (PAW) method was employed with a plane wave energy
cutoff set to 500~eV~\cite{Blochl}. In calculations, we used a 288-atom
hexagonal supercell (equivalent to a $6\times6\times2$ arrangement of the
primitive cell), sampling the Brillouin zone at the $\Gamma$ point only. Ionic
relaxation was conducted until forces fell below 0.01~eV/\AA.

% Ab initio calculations were performed using spin-polariozed density functional
% theory as implemented in VASP~\cite{Kresse_vasp}. The projector augmented wave
% (PAW) method was employed with a plane wave energy cutoff set to
% 500~eV~\cite{Blochl}. The calculations were carried out utilizing a 288-atom
% $6\times 6\times 2$ supercell, and the Brillouin zone was sampled only at the
% $\Gamma$ point. Ionic relaxation was performed until all Hellman--Feynman forces
% dropped below 0.01~eV/\AA.

The formation energies of the defect (refer to Sec.~\ref{sec:formation}) were
calculated using the Heyd--Scuseria--Ernzerhof (HSE) hybrid density
functional~\cite{Heyd}. To accurately capture the material's band gap, the
mixing parameter $\alpha$ was adjusted from the default value of 0.25 to 0.31,
achieving a band gap of 5.95 eV. This computational setup yielded
lattice constants $a = 2.490$~\AA\ and $c = 6.558$~\AA, which show good
agreement with experimental measurements ($a=2.506$~\AA\ and $c=6.603$~\AA ~\cite{Paszkowicz}). Furthermore, the formation enthalpy of the compound was calculated to
be $-2.89$ eV, aligning well with the experimental value of $-2.60$
eV~\cite{Tomaszkiewicz}.

% The formation energies of the defect (see Sec. \ref{sec:formation}) were
% investigated with the Heyd--Scuseria--Ernzerhof (HSE) hybrid density
% functional~\cite{Heyd}, with a mixing parameter $\alpha$ slightly increased from
% the standard value of 0.25 to 0.31 to yield the correct band gap of 5.95~eV. The
% calculated lattice constants of $a=2.490$~\AA\ and $c=6.558$~\AA\ closely match
% experimental values. Additionally, the calculated formation enthalpy of the
% compound, $-2.89$~eV, aligns well with the experimental value of
% $-2.60$~eV~\cite{Tomaszkiewicz}.

The formation energy $E_f$ of a carbon chain tetramer is expressed
as~\cite{Zhang_Northrup,Freysoldt_RevModPhys}
\begin{align*}
  E_f(q) = & E_D(q) - E_H - 4 \mC + 2(\mB + \mN)
  \\
  & + q (E_V + E_F) + E_{\text{corr}}
% \label{eq:formation}
\end{align*}
where $E_D$ and $E_H$ represent the total energies of the supercell with and
without a tetramer, respectively. Additionally, $\mB$, $\mN$, and $\mC$ denote
the chemical potentials of boron, nitrogen, and carbon, respectively. When
calculating the formation energy, we assumed the chemical potential of carbon to
be equal to that in bulk graphite. This represents the maximum possible value of
$\mC$, indicating the highest carbon availability during growth. It is worth
noting that $\mB+\mN=\mu_{\mathrm{BN}}=E_H/n$, where $n$ represents the number
of formula units in the supercell ($n=144$ in our case). Therefore, $E_f$ does
not depend on the individual chemical potentials of B or N. $q$ is the charge
state of the tetramer, $E_V$ stands for the energy of the valence band maximum
(VBM), and $E_{\text{corr}}$ is a correction that describes the interaction
between charged defects in periodic supercells. $E_{\text{corr}}$ value was
calculated using the Freysoldt--Neugebauer--Van de Walle
methodology~\cite{Freysoldt_PRL}. $E_F$ is the Fermi level with respect to
$E_V$. The charge-state transition level $\varepsilon (q/q')$ is defined as the
value of $E_F$ for which the formation energies of two charge states, $q$ and
$q'$, are equal.

In modeling electronic \change[structure]{properties}(Sec.~\ref{sec:electronic}), we utilized the meta-GGA r$^2$SCAN exchange--correlation functional~\mbox{\cite{Sun_SCAN,Furness_r2SCAN}}, which provided a band gap of 4.83~eV. The optimized lattice constants, $a=2.502$~\AA\ and $c=6.515$~\AA, along with the formation enthalpy of $-2.89$~eV (equal to the HSE result), exhibit good alignment with experimental findings, demonstrating the effectiveness of r$^2$SCAN in simulating bulk properties despite underestimation of the band gap.

Our decision to utilize the r$^2$SCAN functional stemmed from the HSE functional's tendency to overestimate ZPL energies. While r$^2$SCAN tends to slightly underestimate ZPL energies, its performance in the context of diamond defects, for instance, yields mean absolute errors for ZPL similar to those obtained using HSE with a standard mixing parameter ($\alpha=0.25$)~\cite{Maciaszek_JCP}. Elevating $\alpha$ to match the experimental band gap implies a potential widening of HSE's deviation in ZPL energy predictions. Furthermore, such adjustment of $\alpha$ raises concerns regarding the accuracy of calculations, as this approach may not satisfy Koopman's theorem~\cite{Ganyecz}. However, for a comprehensive evaluation, all calculations were also performed using HSE with $\alpha=0.31$, and the comparison of results obtained with both methodologies is provided in the Supplementary Material.

% Modeling of the electronic excitations was conducted using meta-GGA r$^2$SCAN% exchange--correlation functional~\cite{Sun_SCAN,Furness_r2SCAN}. It resulted in the band gap of 4.83~eV. Optimized lattice constants are $a=2.502$~\AA, and $c=6.515$~\AA{} and the formation enthalpy of the compound equals the HSE value, $-2.89$~eV. Therefore, besides the band gap, bulk properties calculated with r$^2$SCAN exhibit very good agreement with experimental values. The rationale for employing additional methodology is as follows: it is known that HSE usually overestimates the ZPL energy even when assuming the standard value of $\alpha$, 0.25. On the other hand, r$^2$SCAN typically slightly underestimates ZPL, and the mean absolute error for ZPL for defects in diamond is comparable for r$^2$SCAN and HSE ($\alpha=0.25$)~\cite{Maciaszek_JCP}. Assuming an increased value of $\alpha$ to yield the correct bandgap, the error of HSE in calculating the ZPL can be expected to be even larger. Moreover, adjusting $\alpha$ to match the bandgap may be questioned, as this approach often leads to the total energy of the system failing to satisfy Koopman's theorem~\cite{Ganyecz}. For the sake of completeness, we performed all the calculations with HSE($\alpha=0.31$) as well, and a comparison of the results obtained with both methodologies is provided in the Supplementary Material.

Van der Waals interactions were taken into account by employing the Grimme-D3
method in HSE calculations~\cite{Grimme} and the nonlocal rVV10
method~\cite{Dion,Roman-Perez,Klimes,Peng_rVV10} in r$^2$SCAN calculations.

\section{Results and discussion}

\subsection{Formation and temperature stability}\label{sec:formation}

\begin{figure}
\includegraphics[width=1\linewidth]{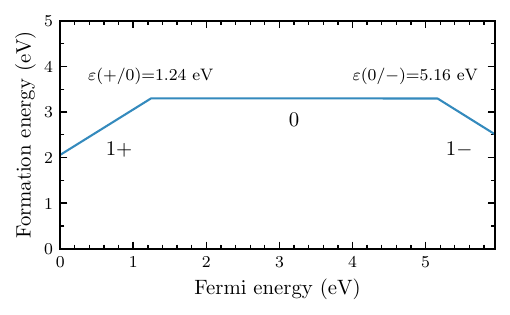}
\caption{Formation energy of \textit{trans}-$\CBCNt$ as a function of the Fermi
  level.\label{fig:formation_energy}}
\vspace{1em}
\end{figure}

The formation energy of \change[the carbon]{the neutral carbon} chain tetramer
(\textit{trans} conformer) is 3.30~eV (Fig.~\ref{fig:formation_energy}). Its
neutral charge state is stable in a wide range of the Fermi level: charge-state
transition levels are 1.24 and 5.16~eV above the VBM for $(+/0)$ and $(0/-)$,
respectively.
% \textcolor{red}{A slight asymmetry between the transition levels
%   with respect to the band edges can be understood when considering the
%   transition levels of isolated $\CB$ and $\CN$ defects. The $\varepsilon(0/-)$
%   of $\CN$ is 2.8~eV, while the $\varepsilon(+/0)$ of $\CB$ is 3.7~eV above the
%   VBM~\cite{Weston,Maciaszek_PRM}. Therefore, the range of the stability of the
%   neutral state of a system of four isolated C monomers (which can form a
%   tetramer) is also slightly shifted towards the conduction
%   band.[\textcolor{blue}{I don't understand this}]}
The $2+$ and $2-$ charge states were found not to be stable.

Regarding the \textit{cis} conformer, its formation energy is lower by 0.06 eV than that of the \textit{trans} conformer, and its charge-state transition levels are $\varepsilon(+/0)=1.07$~eV and $\varepsilon(0/-)=5.23$~eV above the VBM. Since further analysis of the formation and temperature stability focuses on the neutral charge state, where the formation energies are nearly identical for both conformers, the conclusions are applicable to the \textit{cis} conformer as well. Quantitative results pertain to the \textit{trans} conformer.

Although the formation energy of carbon tetramers is not particularly high, blue emitters are not formed under equilibrium conditions but through electron irradiation treatment, and the number of fabricated emitters depends on the electron dose~\cite{Fournier,Gale}. It was demonstrated that the formation of blue emitters is most efficient in samples exhibiting a high concentration of 4.1~eV emitters~\cite{Gale}. Since the 4.1~eV emitters are commonly identified with carbon dimers $\CB\CN$~\cite{Mackoit_dimer}, this experimental observation suggests that blue emitters are carbon related defects. Furthermore, it was speculated that blue emitters are formed through the restructuring of carbon dimers induced by the electron beam, which breaks the bonds and enhances the drift of ionized atoms~\cite{Gale}. This aligns well with our hypothesis, as a carbon tetramer can be formed from two dimers. Alternatively, one could posit that carbon tetramers are not formed by merging two dimers but rather by the fragmentation of larger carbon clusters induced by the electron beam. The presence of such clusters should correlate with the large concentration of carbon dimers, as both phenomena are caused by high carbon density in the h-BN sample.\looseness=1

The blue emitters remain stable up to 800\si{\celsius}. However, after annealing at 1000\si{\celsius} for 1 hour, it was demonstrated that the blue emitters disappear, and other emitters emerge, exhibiting various ZPLs, mostly distributed between 2.58 and 2.95~eV~\cite{Chen_annealing}. In the following, we discuss the results of this experiment in the context of the tetramer model.

Experimental observations suggest that annealing at 1000\si{\celsius} sufficiently enhances diffusion to restore thermodynamic equilibrium within the timeframe of the experiment. The existence of tetramers under thermodynamic equilibrium at 1000\si{\celsius} is unlikely due to their high formation energy. Even assuming the maximal value of the carbon chemical potential, the density of tetramers at 1000\si{\celsius} is $3 \times 10^{10}~\text{cm}^{-3}$. Assuming a flake thickness of 50~nm, this density implies one tetramer per square with a side length of 25~$\mu$m, exceeding the area examined in~\cite{Chen_annealing}. We propose that the disappearance of blue emitters is caused by the dissociation of tetramers, which can be conceptualized as spontaneous fragmentation induced by high temperature. The presence of smaller carbon defects is more likely due to their lower formation energies.

When one of the bonds in a tetramer breaks, smaller carbon complexes such as
dimers, trimers, or monomers are formed. The following are three possible
scenarios of the dissociation of a tetramer:
\begin{align*}
  \CBCNt &\rightarrow 2 (\CBCN)&   \quad (\Delta E=0.99~\text{eV}),
  \\
  \CBCNt &\rightarrow (\mathrm{C}_2 \CN)^{-} + \CB^{+}& \quad (\Delta E=2.71~\text{eV}),
  \\
  \CBCNt &\rightarrow (\mathrm{C}_2 \CB)^{+} + \CN^{-}& \quad (\Delta E=2.87~\text{eV}).
\end{align*}
Here, $\Delta E$ describes the energetic cost of such a reaction, calculated as the difference in formation energies between substrates and products. These values were estimated using theoretical data provided in~\cite{Maciaszek_PRM}. According to the $\Delta$E values, the most probable scenario appears to be the dissociation into two dimers. However, the $\Delta$E values for the latter two reactions correspond to two isolated products (i.e., a monomer and a trimer at a large distance). In actual reactions, the monomer and trimer are in close proximity to each other just after dissociation. Consequently, strong electrostatic attraction reduces $\Delta$E, and this reduction can be estimated to be as significant as 1~eV, or even slightly higher, for certain configurations. Thus, the formation of monomers and trimers is also likely to occur.

The observed diversity in ZPLs can also be attributed to differences in distances between charged carbon defects, such as monomers and trimers, resulting in variations in electrostatic interactions. ZPLs of $C_B^+$ and $C_N^-$ pairs exhibit a strong dependence on the distance between carbon atoms, as demonstrated in~\cite{Auburger}. The similar phenomenon of the emergence of other emitters after annealing was also noted for carbon-based visible (red) emitters, and the optical properties of the produced emitter families are similar in both cases~\cite{Chen_annealing}. This observation aligns well with our identification.

The annealing temperature allows for the estimation of the activation energy of the rate-limiting process. Assuming a typical phonon frequency in h-BN of $10^{14}~\text{s}^{-1}$~\cite{Geick} and a process rate on the order of $1~\text{s}^{-1}$, we can approximate its value to be between 3.0 and 3.5~eV. This value is similar to the migration barrier of boron vacancies in h-BN, which is 3.09~eV for $V_{\mathrm{B}}^{-}$~\cite{Weston}. It can be expected that electron irradiation creates vacancies at a high density. If these vacancies are present close to the tetramer, they can mediate carbon diffusion. However, due to their high formation energies, boron vacancies eventually vanish. It can be hypothesized that since boron vacancies are produced in the vicinity of irradiated spots and thus close to blue emitters, they are able to mediate carbon diffusion; however, after some time, they are annealed.

\subsection{Electronic structure and optical properties}
\label{sec:electronic}

\begin{figure}
  \centering
  \includegraphics[width=0.5\textwidth]{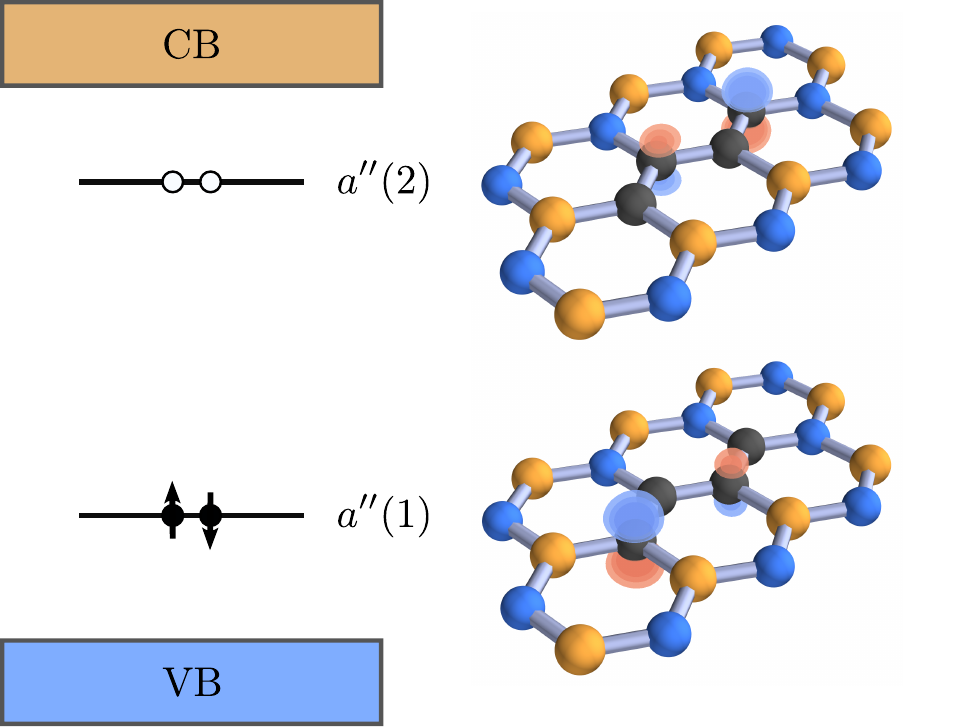}
  \caption{Kohn--Sham electronic energy level diagram and corresponding orbital
    isosurfaces for the \textit{trans} conformer.\label{fig:orbitals}}
\end{figure}

% In the subsequent analysis, the neutral charge state of the tetramer will be
% considered. The electronic structure of the tetramer comprises two levels within
% the bandgap: a lower-lying, fully occupied level, and an upper-lying, entirely
% empty level (Fig. 3). The electronic structure of the blue emitter proposed
% in~\cite{Zhigulin1}, deduced from the photostability of the emission, closely
% resembles the structure of the tetramer presented here.

In the analysis that follows, we focus on the neutral charge state of the tetramer, whose electronic structure features two distinct levels within the bandgap: a lower-lying, fully occupied level, and an upper-lying, vacant level, as shown in Fig.~\ref{fig:orbitals}. The electronic structure of the blue emitter proposed in~\cite{Zhigulin1}, deduced from the photostability of
the emission, closely resembles the structure of the tetramer presented here.

% Both conformers of the carbon chain tetramer exhibit $C_s$ symmetry. Both states
% localized in the bandgap can be labeled according to the irreducible
% represtantation $a'$. For clarity, the lower state will be denoted as $a'(1)$,
% and the upper state as $a'(2)$.

Our computations demonstrate that both conformers of the carbon chain tetramer present $C_s$ symmetry. The localized states within the band gap belong to the irreducible representation $a''$, with the lower-energy state denoted as $a''(1)$ and the higher-energy state as $a''(2)$ (see Fig.~\ref{fig:orbitals}). This configuration indicates that all bound electronic states of neutral charge state exhibit $A'$ symmetry, implying that both permanent (diagonal) and transition (off-diagonal) dipole moments align within the two-dimensional plane of the h-BN lattice.

% The ground singlet state $A'$ can be expressed as $| a'(1) \bar {a'}(1) \rangle$
% (symbols without and with a bar refer to spin-up and spin-down electrons,
% respectively). The system can be excited to either a singlet (S) or triplet (T)
% state. The triplet excited state, $m_s=1$, is represented as
% $| a'(1) a'(2) \rangle$. The singlet excited state is
% $\frac{1}{\sqrt{2}} (| a'(1)\bar{a'}(2) \rangle - | a'(2) \bar {a'}(1) \rangle)$.
% Since it cannot be expressed using a single Slater determinant, determining its
% energy is not straightforward. To calculate it, we adopt the methodology
% discussed in~\cite{Mackoit_dimer}:
%  \begin{equation}
%     E(S)=2 E(m)- E(T)
%  \end{equation}

% where $E(m)$ represents the energy of $| a'(1) \bar {a'}(2) \rangle$, which is a mixed spin state ($m$). Its energy was computed using the delta-self-consistent-field ($\Delta$SCF) approach~\cite{Jones}. The geometry of the excited state S was approximated as the relaxed geometry of the triplet excited state T. Since T is the true wavefunction, its geometry can be expected to provide a more accurate approximation than geometry of state $m$.

In the molecular orbital picture, the ground state is a closed-shell singlet
represented by wavefunction $|S_{0}\rangle = | a''(1) \bar{a}''(1) \rangle$,
where the notation without and with a bar distinguishes spin-up from spin-down
molecular orbitals. The transition to singly excited singlet $|S_{1}\rangle$ and
triplet $|T\rangle$ states involves excitation from $a''(1)$ to $a''(2)$. For
$m_{s}=1$ projection, the triplet state is depicted by a single determinant
wavefunction $|T\rangle = | a''(1) a''(2) \rangle$. The excited singlet state,
$|S_{1}\rangle$, on the other hand, emerges as a strongly-correlated,
multi-determinant state:
$|S_{1}\rangle = 1/\sqrt{2} (| a''(1) \bar{a}''(2) \rangle - | a''(2) \bar{a}''(1) \rangle)$.
Due to its multi-determinant nature, traditional DFT cannot directly model $|S_{1}\rangle$. Thus, we determine its
energy using an approximate spin-purification
approach~\cite{ziegler1977calculation,vonBarth1979}. It involves two distinct
energy computations: one for the single-determinant triplet state and another
for a mixed spin determinant, $|M\rangle = {| a''(1) \bar{a}''(2) \rangle}$. The
energy of the excited singlet state is approximately given
by~\cite{Mackoit_dimer}:
\begin{equation}
  E(S_{1}) = 2E(M) - E(T).
\end{equation}
We employed the delta-self-consistent-field ($\Delta$SCF) methodology,
constraining occupations of Kohn--Sham orbitals, for computing
single-determinant energy terms $E(M)$ and $E(T)$. Concerning the geometric structure, given that
the excited singlet and triplet states possess identical orbital configurations
and differ only due to exchange splitting, it indicates that the geometry of the
excited state $S_{1}$ is close to that of the triplet state. Therefore, we
approximated the geometry of $S_{1}$ by the relaxed geometry of the $T$
state.

The ZPL energies have been calculated as 2.77~eV for the
\textit{trans} conformer and 3.10~eV for the \textit{cis} conformer. Since
the r$^2$SCAN method tends to slightly underestimate ZPL energies, the
calculated value for the \textit{trans} conformer closely matches the
experimental result of 2.85~eV. Notably, using the relaxed geometry of the mixed
spin state ($M$ state) to represent the singlet excited state $S_1$ leads to ZPL
energies lower by about 0.04~eV (all results are detailed in the Supplemental Material). This observation is consistent with the
above assumption that the geometry of the triplet state closely approximates
that of the excited singlet state.

% Relaxation from the singlet excited state S can occur via two processes:
% radiative transition to the singlet ground state or intersystem crossing to the
% triplet state T, which is situated 0.35 eV below the singlet excited state S.
% \textcolor{red}{[A few sentences about ISC - discussing the probability of
%   spin-orbit coupling and hyperfine coupling]}. The rate of the radiative
% lifetime $\tau_{rad}$ is expressed as

% \begin{equation}
% \begin{split}
% \frac{1}{\tau_{rad}} = \frac{n_r E_{\text{ZPL}}^3 \mu^2}{3 \pi \varepsilon_0 \hbar^4 c^3}
% \end{split}
% \end{equation}

% where $n_r$ is the refractive index of h-BN, $E_{\text{ZPL}}$ is the energy of
% the ZPL, $\mu$ is the optical transition dipole moment, and $\varepsilon_0$ is
% the vacuum permittivity. To determine $\mu$, it is necessary to calculate the
% dipole moment between the excited and ground singlet states:
% \begin{equation}
% \begin{split}
% \mu= \langle a'(1) \bar {a'} (1) | \vec {r} | {\frac{1}{\sqrt{2}}} (a'(1) \bar {a'}(2) - \bar {a'} (1) a'(2) )\rangle \\
% = \sqrt{2} \langle a'(1) | \vec{r} | a'(2) \rangle
% \end{split}
% \end{equation}

% Assuming the experimental ZPL ($E_{\text{ZPL}}=2.85~\text{eV}$) and
% $n_r=2.25$~\cite{Lee, Grudinin}, the calculated lifetime for the \textit{trans}
% conformer $\tau_{rad}$ is 1.6 ns, in very good agreement with the experimental
% values of 1.8-2.1 ns ~\cite{Fournier, Gale}. The calculated transition dipole
% moment is $1.77$ e~\AA. Regarding the \textit{cis} conformer,
% $\tau_{rad}=3.6 ns$ and $\mu=1.18$ e~\AA.

Electronic relaxation from the singlet excited state $S_{1}$ can proceed through
two pathways: a radiative transition to the singlet ground state or intersystem
crossing to the triplet state $T$, located 0.35~eV below $S_{1}$. Given the
relatively large exchange splitting of 0.35~eV between the singlet and triplet
states, coupled with minor differences in potential energy surfaces (leading to
small Huang--Rhys factor), it is expected that the non-radiative lifetime
substantially exceeds the radiative lifetime, thereby making radiative decay the
predominant mechanism.

The radiative lifetime, $\tau_{\text{rad}}$, can be estimated using equation:
\begin{equation}
\frac{1}{\tau_{\text{rad}}} = \frac{n_r E_{\text{ZPL}}^3 \mu^2}{3 \pi \varepsilon_0 \hbar^4 c^3},
\end{equation}
where $n_r$ is the refractive index of h-BN, $E_{\text{ZPL}}$ is the ZPL energy,
$\mu$ is the optical transition dipole moment, and $\varepsilon_0$ is the vacuum
permittivity. The transition dipole moment, $\mu$, is estimated utilizing the
expressions derived for molecular orbital states:
\begin{equation}
\mu = \langle S_{0} | \sum_{i} \mathbf{r}_{i} | S_{1} \rangle = \sqrt{2} \langle a''(1) | \mathbf{r} | a''(2) \rangle,
\end{equation}
where the final single-particle expression is obtained by applying the Condon
rules for determinantal states. We then calculate
$\langle a''(1) | \mathbf{r} | a''(2) \rangle$ using single-particle Kohn--Sham
orbitals determined from the triplet electronic and geometric configuration, a
common practice in DFT calculations.

With the ZPL energy set to an experimental value of 2.85 eV and using $n_r = 2.25$~\cite{Lee, Grudinin},
the radiative lifetime for the \textit{trans} conformer is
calculated to be 1.6 ns. This finding aligns closely with experimental lifetimes
reported between 1.8 and 2.1 ns~\cite{Fournier, Gale}. The calculated transition
dipole moment is 1.77 e\AA. For the \textit{cis} conformer, $\tau_{\text{rad}}$
is 2.5 ns, with a transition dipole moment of 1.42~e\AA.

% To estimate the strength of electron-phonon coupling, the Huang-Rhys factor $S$ was calculated within an effective mode approximation~\cite{Alkauskas_PRL}. The $S$ value represents the average number of phonons emitted during the transition. The energy released due to lattice relaxation after the transition to the ground state is 0.20 eV. The calculated effective phonon frequency is $\hbar \omega = 134 meV$, and the Huang-Rhys factor $S$ is 1.50. The experimental value of $S$ can be estimated based on the measured fraction of light emission concentrated on the ZPL. This fraction, called the Debye--Waller factor $DWF$, enables easy calculation of $S$ using the relation $DWF=e^{-S}$. Experimental $DWF$ values are 0.26 in~\cite{Shevitski} and 0.40 in~\cite{Fournier}. Therefore, the experimental Huang-Rhys factors are 1.35 and 0.92, respectively. A discrepancy between calculated and experimental values can be partially attributed to the inaccuracy of the effective mode approximation, which was developed primarily for defects with larger $S$ in solids of low polarity.

To quantify the electron-phonon coupling strength, we calculated the Huang--Rhys
factor $S$ within the effective mode approximation, following the methodology
outlined in Ref.~\cite{Alkauskas_PRL}. The Huang--Rhys factor, $S$,
characterizes the mean number of phonons emitted during electronic transitions.
The energy released due to lattice relaxation after transitioning to the ground
state was calculated to be 0.20~eV. We determined the effective phonon frequency
to be $\hbar \omega = 134~\text{meV}$, resulting in a Huang--Rhys factor of
$S = 1.50$. This factor is pivotal for estimating of the luminescence intensity
fraction within the zero-phonon line (ZPL), referred to as the Debye--Waller
factor (DWF). Employing the single-mode approximation allows for an estimation
of $\mathrm{DWF}\approx e^{-S}$, an approximation applicable in the limit
$\hbar\omega/E_{\text{ZPL}} \ll 1$. This yields a theoretical Debye--Waller
factor of $\mathrm{DWF}=0.22$. Using a more rigorous formula, as explained in
the Supplementary Material, results in a marginally increased value of 0.28. The
experimental measurements of the DWF, which were reported as 0.26 in
Ref.~\cite{Shevitski} and 0.40 in Ref.~\cite{Fournier}, place our calculated
value in agreement with the experimental findings. Regarding the \textit{cis} conformer, the calculated values are $S = 1.91$ and $\hbar \omega = 122~\text{meV}$.

HSE results confirm the general conclusions drawn from r$^2$SCAN calculations.
HSE with $\alpha=0.31$ yields higher values of the ZPL, 3.29~eV and 3.58~eV for
the \textit{trans} and \textit{cis} conformers, respectively. Since HSE
typically overestimates ZPL, especially with an assumed $\alpha$ larger than its
standard value of 0.25, HSE values can be considered upper limits for the ZPL.\@
Consequently, r$^2$SCAN provides lower limits. The experimental value of 2.85~eV
falls within this estimated range for the \textit{trans} conformer. Concerning the radiative lifetime, its values are slightly smaller when calculated using HSE (1.4 ns and 2.2 ns for \textit{trans} and \textit{cis}, respectively). All results obtained with both functionals are presented
in the Supplementary Material. Additionally, all data are calculated for both
approximations of the geometry of the singlet excited state: by the geometry of
the triplet ($T$) and the mixed spin ($M$) states.

% Finally, experimental results concerning the Stark effect will be discussed. It
% was found that an out-of-plane electric field induces only a very weak linear
% Stark shift, with a sensitivity estimated to be 0.04 meV/V nm$^{-1}$. No
% quadratic dependence was observed for the out-of-plane direction. Regarding the
% in-plane electric field, a significant quadratic shift of the ZPL was observed,
% although its linear component is very small~\cite{Zhigulin2}.

Finally, we examine results related to the Stark effect. Experiments revealed
that an out-of-plane electric field causes a negligible linear Stark shift,
estimated at 0.04~meV/V~nm$^{-1}$, and no quadratic dependency was detected in
this direction. Conversely, a significant quadratic shift in the ZPL was noted
under in-plane electric fields, with the linear component remaining very
small~\cite{Zhigulin2}.

It is often assumed that the absence of a linear Stark effect implies the defect
has a center of inversion symmetry. However, this assertion should be correctly
understood in reverse: the presence of an inversion symmetry center in a defect
guarantees the absence of a linear Stark effect. Nevertheless, there are instances where a defect lacking inversion symmetry does not exhibit a linear Stark effect or demonstrates a very faint one. For instance, this phenomenon may occur when the electric field induces identical shifts in both optically active energy levels~\cite{Udvarhelyi}.

As previously mentioned, theory predicts no linear Stark shift for an
out-of-plane field according to group theory. The experimentally observed
minimal linear Stark shift in the out-of-plane direction could be attributed to
sample inhomogeneity.

For the in-plane electric field, we introduce a straightforward \textit{ab
  initio} methodology to evaluate the linear Stark shift. Within the
first-order perturbation theory, the ZPL shift is given by:
\begin{equation}
  \Delta E_{\text{ZPL}} = -\mathbf{E} e
  \left(
    \langle S_{1} | \sum_{i} \mathbf{r}_{i} | S_{1} \rangle
    -
    \langle S_{0} | \sum_{i} \mathbf{r}_{i} | S_{0} \rangle
  \right),
\end{equation}
where $\mathbf{E}$ denotes the applied electric field. Leveraging molecular
orbital wavefunctions, this shift can be expressed using single-particle matrix
elements:
\begin{equation}
  \Delta E_{\text{ZPL}} = -\mathbf{E} e
  \left(
    \langle a''(2) | \mathbf{r} | a''(2) \rangle
    -
    \langle a''(1) | \mathbf{r} | a''(1) \rangle
  \right),
\end{equation}
which can be calculated from Kohn--Sham single-particle states. The amplitudes of the permanent dipole moments are calculated to be smaller than $10^{-3}~e$\AA\ for both $a''(1)$ and $a''(2)$. Insignificant values of amplitudes can also be inferred from the symmetry of the spatial distribution of the charge density (see Fig.~\ref{fig:orbitals}). The obtained results demonstrate that the linear Stark shift for the \textit{trans} tetramer under in-plane electric fields should be negligibly small, which is consistent with experimental observations. A similar understanding can be applied to the \textit{cis} conformer, for which the vanishing of the linear Stark shift is also expected.

\section{Conclusions}

In summary, we presented the results of \textit{ab initio} modeling of the carbon chain tetramer in hexagonal boron nitride. Its calculated optical properties match very well with experimental characteristics of the blue emitter. The calculated ZPL of 2.77~eV closely aligns with the experimental value of 2.85~eV, and the radiative lifetime of 1.6 ns is in good agreement with the measured values of 1.8--2.1~ns. Moreover, we demonstrated that the electron-phonon coupling is relatively weak ($S=1.5$), and the calculated Debye--Waller factor is 0.28, in good accordance with the experimental results (0.26 and 0.40). The linear Stark effect should be absent or extremely weak for both in-plane and out-of-plane external fields, consistent with experimental observations. 

Additionally, we discussed experimental studies of the formation and annealing of blue emitters, proposing that the carbon chain tetramer provides a coherent explanation for their observations. Specifically, our identification of the blue emitter as the carbon tetramer elucidates why the fabrication of blue emitters is efficient only in h-BN samples containing numerous UV emitters. Furthermore, the postulated dissociation of the tetramer with temperature provides an explanation for the observed formation of other emitters with similar ZPLs after annealing at 1000\si{\celsius}, attributing it to the emergence of carbon monomers, dimers and trimers at different distances.

\section*{Acknowledgments}

Research was funded by the Warsaw University of Technology within the Excellence Initiative: Research University (IDUB) programme. Computational resources were provided by the Interdisciplinary Center for Mathematical and Computational Modelling (ICM), University of Warsaw (Grant No. GB81-6), and the High Performance Computing center “HPC Saul{\.e}tekis” in the Faculty of Physics, Vilnius University.

\bibliographystyle{iopart-num}
\bibliography{bib.tex}

\end{document}